\documentstyle[twoside,fleqn,espcrc22]{article}
\newcommand{\ba}{\begin{eqnarray}}  
\newcommand{\ea}{\end{eqnarray}}  
\newcommand{\be}{\begin{equation}}  
\newcommand{\ee}{\end{equation}}

\newcommand{\re}{\mbox{Re}}  
\newcommand{\im}{\mbox{Im}}  
  
\newcommand{\ugu}{\!\!\!&=&\!\!\!}

\newcommand{\nn}{\nonumber}

\begin{document}
\title{
 Measuring the $a_0-a_2$ pion scattering
lengths through  $K \to 3\pi$ decays 
\thanks{Invited talk given by I.S. at ``QCD 06'', 3-7 July 2006, Montpellier, France.}
}
\author{ Ignazio Scimemi,\address{
 Departament de F\'{\i}sica Te\`orica, IFIC,
 CSIC-Universitat de Val\`encia,\\  
Apt. de Correus 22085, E-46071 Val\`encia, Spain. }
Elvira G\'amiz\address{Department of Physics, University of Illinois, Urbana
IL 61801, USA.}
 and  Joaquim Prades
\address{
Theory Unit, Physics Department,
CERN, CH-1211 Gen\`eve 1211, Switzerland. 
}
\thanks{
On leave of absence from 
CAFPE and  Departamento de F\'\i sica Te\'orica 
y del Cosmos, Universidad de Granada,
Campus de Fuente Nueva, E-18002 Granada, Spain.}  
}

\begin{abstract}  
We discuss the recent Cabibbo's proposal to measure the  
$\pi\pi$ scattering lengths combination $a_0-a_2$ from the    
cusp effect in the $\pi^0\pi^0$  energy spectrum  at threshold   
for $K^+ \to \pi^0 \pi^0 \pi^+$  and $K_L\to \pi^0 \pi^0\pi^0$.
We estimate  the theoretical uncertainty of the $a_0-a_2$ determination  
at NLO in our approach and  obtain that it is not smaller   
than  5 \% 
for $K^+\to\pi^0\pi^0\pi^+$.  One gets  similar theoretical uncertainties
if the neutral  $K_L\to\pi^0\pi^0\pi^0$ decay  data below threshold
are used instead. For this decay, there are very large  
theoretical uncertainties above threshold 
due to cancellations and  data above threshold 
cannot be used to get the scattering lengths. \\

Preprint CERN-PH-TH/2006-219 , IFIC/06-37
\end{abstract}


  \maketitle
\section{Introduction}  
  
  Final state interaction (FSI) phases   
at next-to-leading order (NLO) in Chiral Perturbation Theory for $K\rightarrow 3 \pi$
are an important ingredient to obtain   the charged CP-violating asymmetries at  NLO  
\cite{GPS03,SGP04,PGS05}. 
 The dominant contribution to these $K^+\to 3\pi$ FSI at NLO  
are from two-pion cuts and
they were calculated analytically in \cite{GPS03}.

Though to get the full  
$K\to 3\pi$ amplitudes at order $p^6$  implies a  two-loop calculation,  
 one can get the  FSI phases at NLO  using the optical theorem   
within CHPT   with the advantage that one just needs   
to know  $\pi\pi$ scattering   
and  $K\to 3\pi$ both at ${\cal O}(p^4)$. Notice that  
NLO in the dispersive part of the amplitude   
means one-loop and ${\cal O} (p^4)$  
in CHPT while  NLO  in the absorptive part of the amplitude  
means two-loops and ${\cal O} (p^6)$ in CHPT.

 The study of  FSI in $K\to 3\pi$ at NLO  
 also became of relevance after the proposal by Cabibbo  
\cite{CAB04} to measure  the combination $a_0-a_2$ of $\pi\pi$  
scattering lengths using  the cusp effect in  the   
$\pi^0\pi^0$ spectrum at threshold in   
$K^+ \to \pi^0\pi^0\pi^+$ and $K_L\to\pi^0\pi^0\pi^0$
decay rates.\footnote{The cusp effect in SU(2) $\pi\pi$
scattering was discussed in \cite{MMS97}.}  Within this proposal,   
it has been recently presented in \cite{CI05} the effects of FSI at NLO   
using formulas dictated by unitarity and analyticity   
approximated at second order in the $\pi\pi$ scattering lengths,   
$a_i \sim 0.2$. The error was therefore  
 canonically assumed to be of order of $a_i^2$, i.e., $5\%$.  
 There, they used a second order polynomial in the relevant final
two-pion invariant  energy $s_3$  fitted to data
to describe the 
$K\to 3\pi$ vertex that enters in the formulas of the cusp effect. 
It is of interest to check  this canonical error and provide  
a  complementary analysis  of this theoretical uncertainty
estimate. 

In  Ref.~\cite{ktrepi}, we use our NLO in CHPT results for the real part 
of $K\to 3\pi$ fitted to data to describe  the
$K\to 3\pi$ vertex that enters in formulas of the cusp effect.  
Notice that, as we want to extract the pion scattering lengths, 
we do {\em not} want to predict
the real part of  $K\to 3\pi$ in CHPT at any order 
but to have the best possible description fitted to data.
We treat  $\pi\pi$ scattering near threshold as in   
Cabibbo's original proposal. The advantage of using CHPT formulas
for the fit to data of the real part of $K\to 3\pi$  is that 
it contains the correct singularity structure at NLO in CHPT
which can be systematically  improved by going at higher orders.

Contributions from  next-to-next-to-leading order (NNLO) SU(3) CHPT 
in the isospin limit 
are expected typically to be around $(3\sim5)\%$, so that our  
NLO results are   just a first step in  order to reduce   
the theoretical error on the determination of the combination   
$a_0-a_2$ to  the few per cent level.  
At NNLO,  one can   follow a procedure analogous to the one  we 
use in~\cite{ktrepi}  
to get a more accurate  
measurement of $a_0-a_2$ and check the assumed NNLO uncertainty.  
At that point, and in order to reach the few per cent level in the  
theoretical uncertainty,   
it will be necessary   to include full isospin breaking effects 
at NLO too. These are also expected to be  
of a few per cent as was found for $K\to 3\pi$ in  \cite{BB04,BB05}.  
\subsection{Basic notation for charged Kaon decays}  
\label{charged}  
  Near $\pi^+\pi^-$ threshold, 
we can decompose the $K^+\to\pi^0\pi^0\pi^+$ amplitude 
as follows \cite{CAB04,CI05}  
\ba\label{ABdefinition}  
A_{00+} \!\ugu\! \left\{\!\!
\begin{array}{ll}
 \overline A_{00+} +   
\overline B_{00+}v_{\pm}(s_3), &\!\!\! {\rm for\ }  s_3 > 4m_{\pi^+}^2  
\\
 \overline A_{00+} +   
i\overline B_{00+}v_{\pm}(s_3) , &\!\!\! {\rm for\ } 
 s_3 < 4m_{\pi^+}^2  ,
\end{array} \right. \nonumber \\ &&\!\!\!
\ea  
where $\overline A_{00+}$ and $\overline B_{00+}$ are 
in general singular functions except near the $\pi^+\pi^-$ threshold 
\cite{CGKR06} and 
\be  
v_{ij}(s)=\sqrt{\frac{\vert s-(m_{\pi^{(i)}}+m_{\pi^{(j)}})^2\vert}{s}}.  
\ee  
Notice that these kinematical factors are taken with  physical pion
masses, in this way one can describe the cusp effect which is 
 generated by the different behavior  of $K^+ \to \pi^0 \pi^0 \pi^+$
for the two neutral pions invariant energy  above and below  the  
$s_3= 4 m_{\pi^+}^2$ threshold.
 
With these definitions, the differential decay rate for this amplitude   
can be written as \cite{CI05}  
\ba  
\label{tot}  
 | A_{00+}|^{2} \equiv    
\re \overline A_{00+}^2 + \Delta_{A} + v_{\pm}(s_3)   
\Delta_{\rm cusp} \,,  
\ea
with  
\ba  
 \Delta_{A} \!\!\! &\equiv&\!\!\!
 \im \overline A_{00+}^2 + v^2_{\pm}(s_3)   
\left\lbrack  
\re \overline B_{00+}^2 + \im \overline B_{00+}^2 \right \rbrack \, ,\nn \\
\Delta_{\rm cusp}\!\!\!\! &\equiv & \!\!\!\!
 \left\{   
\begin{array}{ll}\!\!\!    
      -2 \re \overline A_{00+}\im \overline B_{00+} +   
2\im \overline A _{00+}\re \overline B_{00+} \, , \\
  {\rm for} \,\,\,  s_{3}<4m_{\pi^{+}}^{2};&  \\  
      2 \re \overline A_{00+}\re \overline B_{00+} +    
2 \im \overline A_{00+}\im \overline B_{00+}\, , 
 \\  {\rm for} \,\,\,  s_{3} > 4m_{\pi^{+}}^{2} . & \end{array} 
\right.   \nn
\label{Dcusp}  
\ea
The combination of real and imaginary amplitudes $\Delta_{\rm cusp}$   
defined above parametrizes the cusp effect due to the   
$\pi^+\pi^-\to\pi^0\pi^0$   
re-scattering in the $K^+\to\pi^0\pi^0\pi^+$    decay rate.

\section{Discussion}  
\label{SC}  
 Our method  is 
a variation of the original Cabibbo's proposal that uses NLO CHPT
for the real part of $K\to 3\pi$ vertex instead of the
 quadratic polynomial in $s_3$ approximation used in \cite{CAB04,CI05}
plus analyticity and unitarity. 

  Notice that we do not use CHPT to predict
the real part  of $K\to 3 \pi$,  but use its 
exact singularity form at NLO in CHPT to fit it  to data above threshold.
If the two-loop CHPT singularity structure were known 
it could be used in order to take this singularity structure
exactly in $\re \overline A_{00+}$. 
The treatment of $\pi\pi$ scattering near threshold 
is independent of this choice
and we treat it in the same way as in \cite{CI05}.

 The cusp effect originates in the different 
contributions to  $K^+\to \pi^0\pi^0\pi^+$   and
$K_L\to \pi^0\pi^0\pi^0$  
amplitudes above and below threshold of $\pi^+\pi^-$ production
in the $\pi^0\pi^0$ pair invariant energy. 
We obtain these contributions 
using  just analyticity and unitarity, in particular applying
Cutkosky rules and   the optical theorem above and below threshold
to calculate the discontinuity across the physical cut.
  This allows us to separate $\pi\pi$ scattering 
--which we want to measure-- from the rest  of $K^+\to \pi^0\pi^0\pi^+$
or $K_L\to \pi^0\pi^0\pi^0$. 

The validity of the use of Cutkosky
rules for  in $K \to 3\pi$ decays is commented in \cite{ktrepi}.
 In particular, the real part of the discontinuity
 has a singularity when any of the $s_i$ invariant energy
reaches its pseudo-threshold at $(m_K-m_{\pi^{(i)}})^2$ 
as described in \cite{CGKR06,GAS06,ANI03}. 
This  singularity affects 
  the description of the cusp effect using the discontinuity
as dicussed in \cite{ktrepi,CGKR06,GAS06}.

We would like to remark here
that making the same approximations  that were done 
 in \cite{CI05} we fully agree with their analytical results.
In particular, we checked that
the use of the quadratic polynomial in $s_3$ in
\cite{CI05} produce negligible differences 
--around 0.5  \%-- in $\Delta_{\rm cusp}$ in (\ref{Dcusp}).

We also pointed out 
that while the presence of that singularity at pseudo-thresholds
does not affect $\re \overline{B}_{00+}(s_3)$  
and $\re \overline{B}_{000}'(s_3)$
when $s_3$ is around threshold, 
one needs to take fully into account its  effects for   
$\delta \re {\overline A_{00+}}$ 
and $\delta \re {\overline A_{000}'}$ when  $s_1$ or $s_2$ is  
above $(m_K^2-m_\pi^2)/2$.
For a possible solution  of this problem   which does not simply use
the discontinuity to describe the cusp effect see  \cite{CGKR06}.  
Another possibility  could be to  use,  instead
of  $\delta \re {\overline A_{00+}}$  
 and $\delta \re {\overline A_{000}'}$, 
 the full-two loop (not available yet) finite relevant pieces
to describe  the singularities at   thresholds  at NLO 
 in $\re {\overline A_{00+}}$   and $\re {\overline A_{000}'}$, 
respectively. This could be fitted to data.   

In~\cite{ktrepi} we have also discussed the approximations 
done in \cite{CI05} and the  numerical differences they induce 
in $\Delta_{\rm cusp}$. We have 
found that though each one of them  is individually negligible
(between 0.5 \% to 1 \%) they produce final differences in the
$\Delta_{\rm cusp}$ around  3 \%.  Of course, these approximations
can be eliminated.
 
Concerning the theoretical uncertainties  in the determination of 
  $a_0-a_2$ using  our formulas,
we concluded that for $K^+\to \pi^0\pi^0\pi^+$, 
 this uncertainty is  somewhat larger than 
5 \% if uncertainties are added quadratically  and 7 \% if added linearly. 
 I.e., we essentially agree with the estimate in \cite{CI05}.
Notice that we get our final theoretical uncertainty
as the sum of several order 1\% to 2\% uncertainties to the canonical
NNLO 5 \% uncertainty. 

For the  case $K_L\to \pi^0\pi^0\pi^0$, we get
 that --if  one  uses  just  data below threshold--
the uncertainty in the determination of
$a_0-a_2$ is  of the same order  as for
$K^+ \to \pi^0 \pi^0 \pi^+$.  Above threshold,  we found  
large numerical cancellations  which preclude from using it.

An expansion in the scattering lengths $a_i$ and Feynman diagrams  
were  used in \cite{CI05} to do the power counting and  
obtain  the cusp effect description of  $K^+\to \pi^0\pi^0\pi^+$
at NLO.  In general, when FSI $\pi\pi$ scattering effects
are included at $n$-th 
order\footnote{$n=1$ order are for LO contributions.},
 there  appear new topologies in $K\to 3\pi$ 
which give contributions  to $\Delta_{\rm cusp}$ 
of order  $a_i^{n}$. 
 The canonical uncertainty of the $n$-th order results  is thus $a_i^{n}$.
Notice that the velocity factors that 
appear after applying  the unitarity  cuts 
can be order one  --for instance
$v_{\pm}((m_K^2-m_\pi^2)/2)\simeq 0.6$ appear
in Re $\overline{B}_{00+}( 4 m_\pi^2)$--
and  do not suppress largely the naive  $a_i^n$ estimate.

Our estimate  for the uncertainty from NNLO, $\sim a_i^2$,  
coincides numerically with the one made 
in  \cite{CI05}, i.e. it is around 5 \%.
We conclude that one cannot expect to decrease this canonical $5\%$
 theoretical uncertainty  of the NLO result 
unless one includes  $\pi\pi$ scattering effects at NNLO. 
If one wants  to  
reach the per cent level in the uncertainty of the  
determination of $a_0-a_2$ from the cusp effect,   
one would need  to include those  NNLO   re-scattering   effects.  
As said above, at NNLO it is possible to   follow a procedure analogous to 
the one  we use  here   to get a more accurate  
measurement of $a_0-a_2$ and check the estimated NNLO uncertainty.  

We have just included isospin breaking due to the different thresholds 
using   two-pion physical phase spaces in the optical theorem
and Cutkosky rules.
This is needed to describe the cusp effect.
The rest of NLO isospin breaking is expected to be  important  just 
at NNLO. At that order,  isospin breaking effects in $\pi\pi$ 
scattering at threshold --both from quark masses
and from electromagnetism--  will have to be implemented 
and their  uncertainties added.  
  
 Finally, we believe that it is interesting to continue investigating
the proposal in \cite{CAB04,CI05} to measure
the non-perturbative $\pi\pi$ scattering lengths from the cusp
effect in $K^+\to \pi^0 \pi^0\pi^+$ and 
$K_L\to \pi^0 \pi^0\pi^0$. Another interesting direction is to develop
an effective field theory in the scattering lengths which could both
check the results in \cite{CI05} and allow to go to NNLO.
 This type of studies is already underway and firsts results were
presented \cite{CGKR06,GAS06}.

\section*{Acknowledgments}  
This work has been supported in part by the European Commission (EC)  
 RTN Network FLAVIAnet Contract No. MRTN-CT-2006-035482 (J.P. and I.S.),
 by  MEC (Spain)  and FEDER (EC)  Grant Nos. FPA2003-09298-C02-01 (J.P.)  
 and  FPA2004-00996 (I.S.), and by Junta de Andaluc\'{\i}a    
Grants No. FQM-101 (J.P.) and FQM-347 (E.G. and J.P.).
I.S. wants to thank the Departament d'ECM, Facultat de F\'{\i}sica, 
Universitat de Barcelona (Spain) for kind hospitality.


\end{document}